\documentclass[preprint,1p,times]{elsarticle}

\usepackage{amssymb,amsmath,graphicx,color,ulem}
\usepackage{hyperref}

\journal{Journal of Subatomic Particles and Cosmology}

\begin{document}

\begin{frontmatter}

\title{Transverse momentum fluctuations as a probe of thermalization and collective dynamics of QGP}

\author[1]{Rupam Samanta}
\affiliation[1]{Institute of Nuclear Physics, Polish Academy of Sciences,  31-342 Cracow, Poland}
\ead{rupam.samanta@ifj.edu.pl}
\author[2]{Tribhuban Parida}
\affiliation[2]{Department of Physical Sciences,
Indian Institute of Science Education and Research Berhampur, Transit Campus (Govt ITI), Berhampur-760010, Odisha, India}

\begin{abstract} 
We study fluctuations of mean transverse momentum per particle ($[p_T]$) in ultrarelativistic heavy-ion collisions. We show that the steep fall in the variance of transverse momentum fluctuation in ultracentral Pb+Pb collision serves as a natural consequence of the thermalization of the QGP medium. We study the correlation between the spectra and $[p_T]$ which maps these fluctuations differentially and dubbed as $v_0(p_T)$. We highlight the importance of $v_0(p_T)$ showing that it plays similar role as anisotropic flow when probing the collective nature of QGP.    
\end{abstract}

\end{frontmatter}

\section{Introduction}
Ultrarelativistic collision of two nuclei creates a hot fluid-like medium known as the Quark Gluon Plasma (QGP), the dynamics of which can be described by relativistic visocous hydrodynamics and the evidence of which has traditionally relied on the anisotropic flow of the particles driven by the initial pressure gradient~\cite{Heinz:2013th,Busza:2018rrf}. However, the evidence is indirect, as it depends on particles' direction (azimuths). We show how the event-by-event fluctuations of the mean transverse momentum per particle ($[p_T]$) in ultracentral collisions, can serve as a direct probe of the QGP formation (thermalized medium), which solely depend on the magnitude of the particles' momenta instead of their direction of flight. Moreover, we show how the correlation between particle spectra and $[p_T]$ can map this fluctuations ($\sigma_{p_T}$) as a function of $p_T$ through a new observable $v_0(p_T)$.        

\section{Mean transverse momentum fluctuations in ultracentral Pb+Pb collision}
The energy deposited in a heavy-ion collision is shared by the particles produced. This is characterized by the mean transverse momentum per particle, $[p_T] \equiv \sum p_{T_i}/N_{ch}$, where $N_{ch}$ is the total number of charged particles produced. $[p_T]$ fluctuates event-by-event, leading to a net dynamical fluctuations at the end. ATLAS measures the variance of this dynamical fluctuations as a function of $N_{ch}$. The relative dynamical fluctuations are small, around 1 $\%$. Interestingly, the data encounter a sharp decrease over a narrow range of $N_{ch}$ in ultracentral region. We argue this sudden decrease as a natural consequence of thermalization of the QGP medium produced in the collision. 
\begin{figure}[h]
    \centering
    \includegraphics[width=0.35\linewidth]{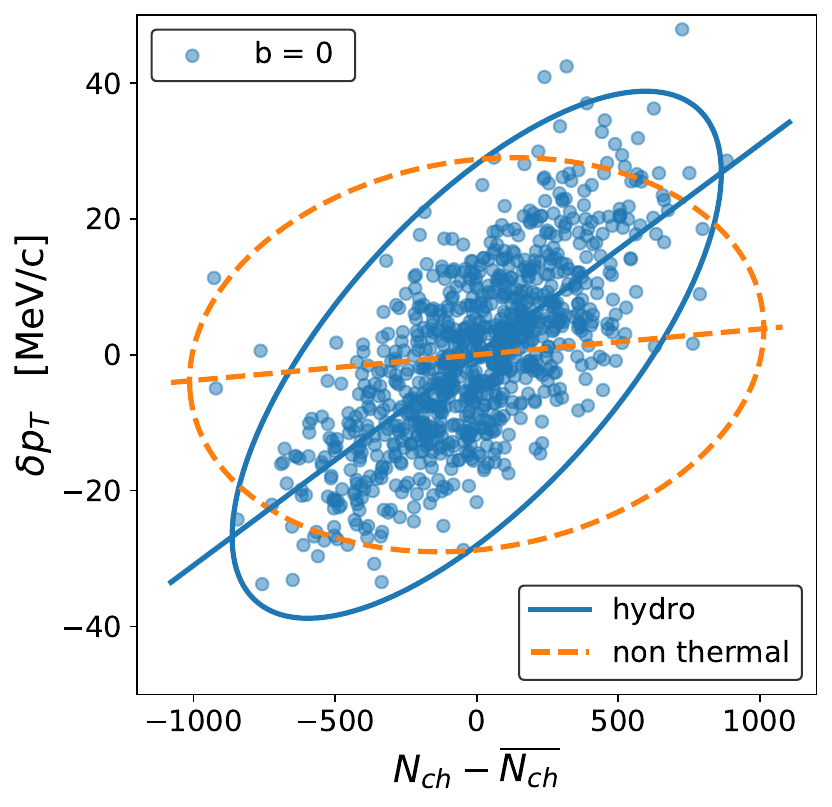}~~~~~~~~~~
    \includegraphics[width=0.4\linewidth]{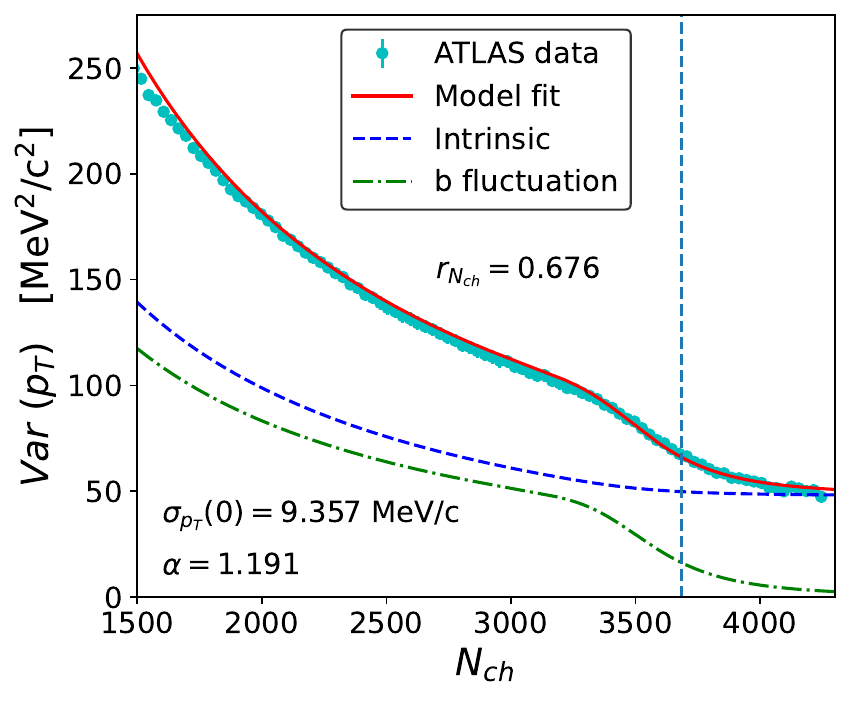}
    \caption{Left: Two dimensional scattered plot of $[p_T]$ and $N_{ch}$ at $b=0$. The blue symbols denote the hydro results and the straight line represents the mean at fixed $N_{ch}$. The elliptic curve correspond to the 99 $\%$ confidence ellipse of the correlated Gaussian distribution. The orange color shows the similar results from a non-thermal HIJING model calculation. Right: Variance of $[p_T]$ as a function of $N_{ch}$ in 20 $\%$ most central Pb+Pb collisions. The blue points and the red lines show the ATLAS measurements and our model fit to the data respectively. The blue dashed curve and the green dashed-dotted curve represent the contribution of intrinsic and impact-parameter fluctuations respectively.}
    \label{fig:sigptvsNch}
\end{figure}
In order to model this dependence, we perform event-by-event hydro simulations of 1000 Pb+Pb collisions at fixed $b$ ($b=0$) as shown in Fig.~\ref{fig:sigptvsNch}. The figure shows that at a constant volume (fixed $b$) a larger number of particle production implies a larger $[p_T]$.  We assume that this distribution can be described by a 2D correlated Gaussian distribution $P([p_T],N_{ch}|b)$ which characterized by five parameters: $\overline{N_{ch}}(b)$, $\sigma_{N_{ch}}(b)$, $\overline{p_T}(b)$, $\sigma_{p_T}(b)$ and $r(b)$, where $r(b)$ denotes the Pearson correlation coefficient between $[p_T]$ and $ N_{ch}$ at fixed $b$~\cite{Samanta:2023amp}. 
The distribution of $ p_T$ (or $\delta p_T \equiv [p_T]-\overline{p_T} $) at fixed $N_{ch}$ is obtained by taking the superposition of the distributions at fixed $b$ and dividing by $P(N_{ch})$ : $P(\delta p_T|N_{ch})= \bigg(\int P(\delta p_T, N_{ch} | b) P(b) db\bigg)/P(N_{ch})$, from which the variance can be obtained by taking the square width,
\begin{equation}
\begin{aligned}
    \text{Var}(p_T|N_{ch}) = \langle (\delta p_T -\langle \delta p_T\rangle)^2\rangle = \text{Var}( p_T|N_{ch},b) + \bigg((\langle \delta p_T^2\rangle_b-\langle \delta p_T\rangle_b^2) | N_{ch} \bigg) \ .
    \label{eq: varpt}
\end{aligned}
\end{equation}

Eq.~\ref{eq: varpt} shows that there are two separate contributions to the variance of $[p_T]$ at fixed $N_{ch}$: i) variance at fixed $N_{ch}$ and fixed $b$ ii) variance of $[p_T]$ due to the fluctuation of $b$ at fixed $N_{ch}$. The first contribution can be understood from Fig.~\ref{fig:sigptvsNch} (left), where even at fixed $b$ (=0) and fixed $N_{ch}$, there are fluctuations of $[p_T]$ which can be attributed to intrinsic fluctuations. The second contribution comes from the fluctuations of impact parameter at fixed multiplicity (see Fig.1 of~\cite{Samanta:2023kfk} and Fig.2 of ~\cite{Samanta:2023amp}). At fixed $N_{ch}$, a larger $b$ corresponds to a smaller collision volume~\cite{Samanta:2023kfk}, hence larger density $N_{ch}/V$. From relativistic thermodynamics, a larger density results in larger energy per particle eventually producing larger $[p_T]$. The distribution of impact parameter at fixed $N_{ch}$ is obtained using Bayes' theorem : $P(b|N_{ch})= \frac{P(N_{ch}|b)P(b)}{P(N_{ch})}$. In Eq.~\ref{eq: varpt} $\langle \dots \rangle$ denotes the mean calculated w.r.t the distribution $P(\delta p_T | N_{ch})$ and $\langle \dots \rangle_b$ denotes the mean calculated w.r.t the distribution $P(b|N_{ch})$.

The variance $\text{Var}(p_T | N_{ch})$ calculated from Eq.~\ref{eq: varpt} is fitted to ATLAS data for 20 $\%$ most central events~\cite{Samanta:2023amp}, shown in Fig.~\ref{fig:sigptvsNch}.  
In the same figure we also show the two separate contributions in Eq.~\ref{eq: varpt}. It could be seen that the contribution of $b$-fluctuations gradually goes to zero in the ultracentral limit, causing the sharp decrease. The values of the fit parameters are shown on Fig.~\ref{fig:sigptvsNch}\footnote{For more details see~\cite{Samanta:2023amp}}. Our model returns a very large value of $r$ implying a very strong correlations between $N_{ch}$ and $[p_T]$ (missing in non-thermal model e.g. HIJING cf. Fig.~\ref{fig:sigptvsNch}) stemming as a direct consequence of the thermalization of the QGP medium.

\section{Mean transverse momentum-spectra correlation : $v_0(p_T)$}
Event-by-event fluctuations of mean transverse momentum per particle result from event-by-event fluctuations of $p_T$-spectra. A softer spectra results in a $[p_T]$ smaller than $\langle [p_T] \rangle$, and a harder spectra gives rise to a larger $ [p_T]$. Therefore, it is intriguing to explore how these two quantities are correlated and what one can learn from that. Following the proposal by Teaney et al.~\cite{Schenke:2020uqq}, we construct this correlation, dubbed as $v_0(p_T)$~\cite{Parida:2024ckk},
\begin{equation}
 v_0(p_T)\equiv\frac{\langle \delta N(p_T)\delta p_T\rangle}{N_0(p_T) \sigma_{p_T}} \ \text{and} \ \ v_0 = \frac{\sigma_{p_T}}{\langle p_T \rangle}
    \label{eq: v0pt}
\end{equation}
where $N(p_T)$ denotes the normalized spectra, $\delta N(p_T)= N(p_T)- \langle N(p_T)\rangle$, $\delta p_T= [p_T]- \langle [p_T]\rangle$, $\langle [p_T] \rangle \equiv \langle p_T \rangle$ and $\langle N(p_T)\rangle = N_0(p_T)$. 
\begin{figure}[h]
    \centering
    \includegraphics[width=0.35\linewidth]{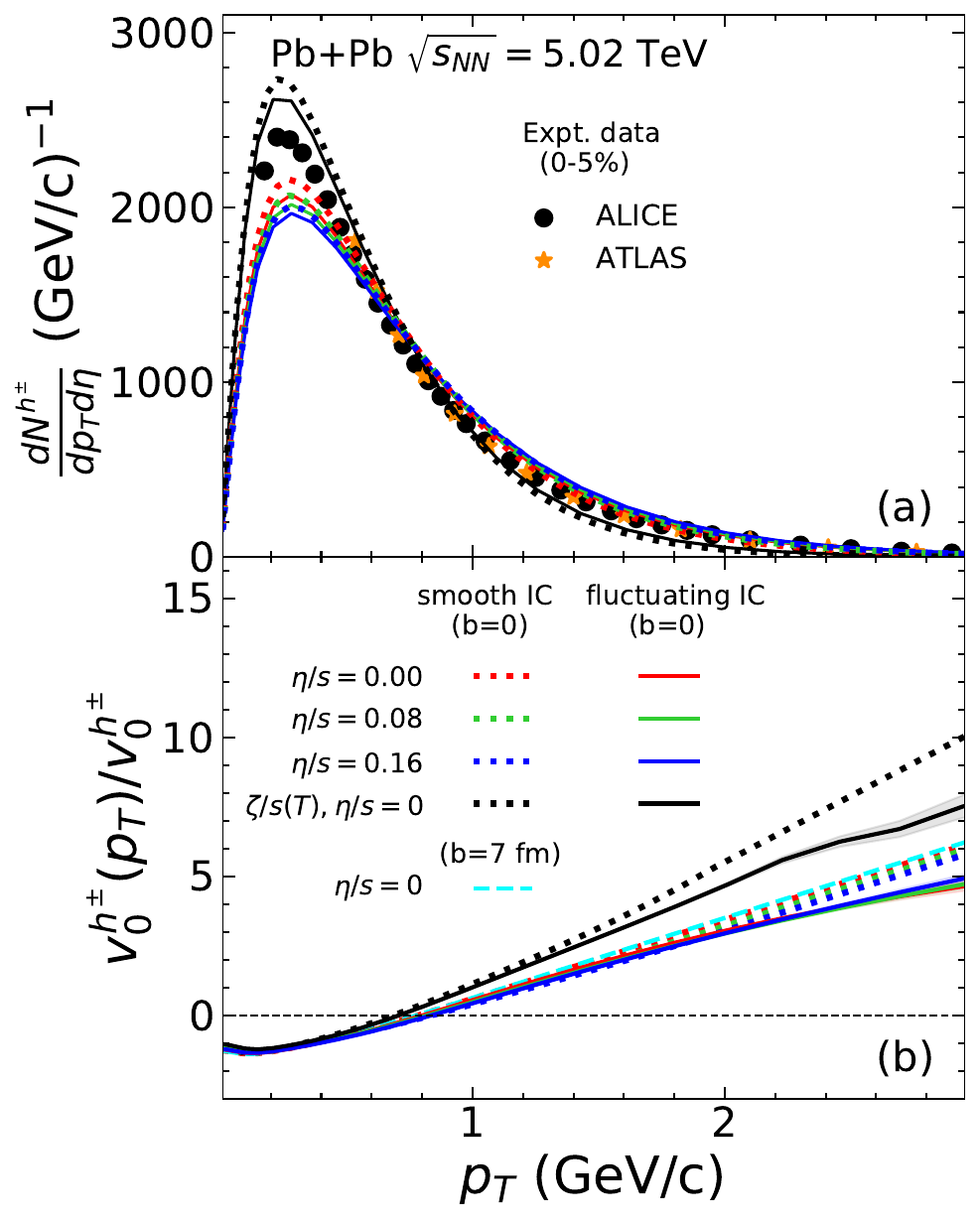}~~~~~~~~~~~
    \includegraphics[width=0.38\linewidth]{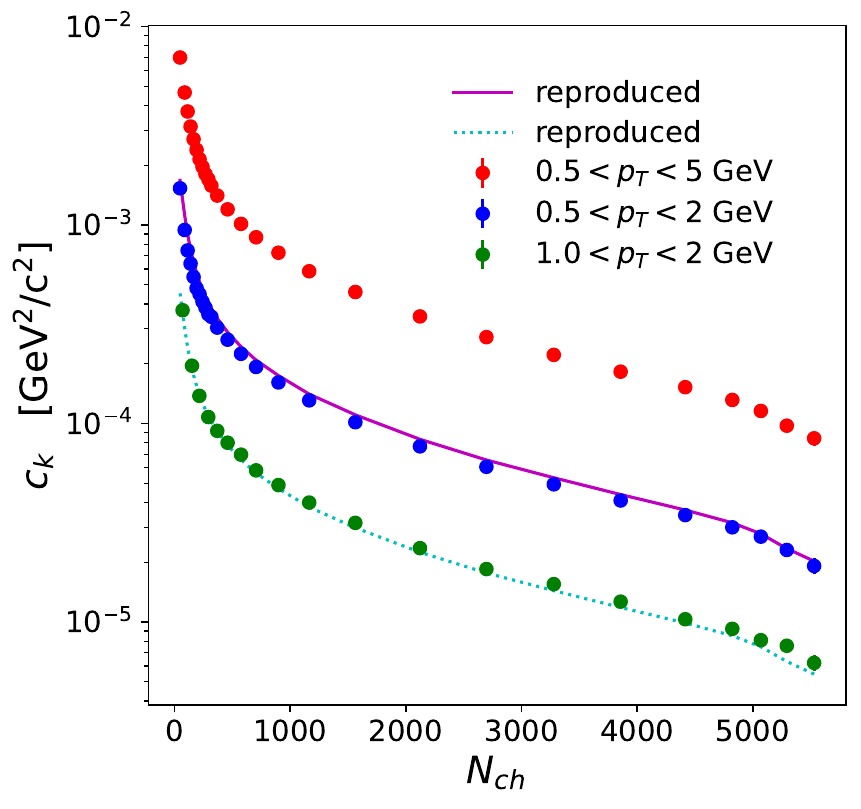}
    \caption{Left: Normalized charged particle spectra (top) and $v_0(p_T)/v_0$ (bottom) as a function of $p_T$ at $b=0$. Results obtained from hydrodynamic simulations are shown for shear and bulk viscosity with smooth and fluctuating initial conditions. Results for $b=7$ fm is also shown as a comparison. Right: Variance of $[p_T]$ for different $p_T$-windows as a function of multiplicity. The circular symbols denote the ATLAS measurements and the solid curves represent the reproduced results in our model taking the data in the largest window as inputs and using the correction factor $C_A$ derived from $v_0(p_T)/v_0$}
    \label{fig:v0pt}
\end{figure}
While $v_0$ represents the total relative fluctuations of $[p_T]$, $v_0(p_T)$ provides a differential estimate of that total fluctuations. The left panel of Fig.~\ref{fig:v0pt} present the results for $v_0(p_T)$ along with the spectra obtained from hydrodynamic simulations at fixed impact parameter. Instead of $v_0(p_T)$, the scaled observables $v_0(p_T)/v_0$ is presented, which is independent of centrality (or $b$), as seen from the figure. Two sets of hydro results are presented: one with fluctuating initial conditions similar to what is done in experiments, and the other one is with the smooth initial conditions where we perform only two hydro simulations. 
In case of hydro with smooth initial conditions, the observable can be calculated as, $\frac{v_0(p_T)}{v_0} \equiv \frac{\delta\ln N(p_T)}{\delta\ln [p_T]}$ discussed in details in ~\cite{Parida:2024ckk}. We consider different scenarios with sheer and bulk viscosity, and find that $v_0(p_T)/v_0$ shows negligible sensitivity to $\eta/s$ while showing a little sensitivity to $\zeta/s$ at large $p_T$. This sensitivity to bulk viscosity is partly due to the sensitivity of mean $\langle p_T \rangle$ to bulk viscosity which is washed out when the horizontal axis is scaled by $\langle p_T \rangle$ i.e. plotting as a function of $p_T/\langle p_T \rangle$~\cite{QMtalkTParida}. 

The observable $v_0(p_T)$ is as important as the anisotropic flow in the sense that it shows similar mass hierarchy as the elliptic flow $v_2(p_T)$ when studied for identified particles~\cite{Parida:2024ckk}, and can be measured in experiments through a similar method~\cite{ALICE:2025iud,ATLAS:2025ztg}. Therefore $v_0(p_T)$ could be termed as the true {\it radial flow}. Another important feature of this observable is that using $v_0(p_T)/v_0$ one could estimate $\langle p_T\rangle$ and $\sigma_{p_T}$ in any $p_T$-acceptance range taking inputs in a particular acceptance, through an acceptance correction factor $C_A$~\cite{Parida:2024ckk,Bhatta:2025oyp}. The right panel of Fig.~\ref{fig:v0pt} shows $c_k \equiv \sigma_{p_T}^2$ as a function of $N$ for three different $p_T$ windows. Taking the red points ($0.5<p_T<5$ GeV) as the inputs and using $C_A$ we reproduce the data at the other two $p_T$ windows: $0.5<p_T<2$ GeV and $1<p_T<2$ GeV. 
\section{Conclusions and Outlook}
We show that the fluctuations of mean transverse momentum per particles can provide direct robust probes of the QGP medium and its collective behavior. The sharp decrease of the variance in ultracentral collisions is caused by the rapid decline in impact parameter fluctuations, providing direct evidence of the underlying thermalization of QGP manifested in hydrodynamics. The observable $v_0(p_T)$ characterizes these fluctuations at differential level and its similarities with anisotropic flow establish it as a true measure of the radial flow which directly probes the collective dynamics of the QGP medium. Extension of these studies to small systems such as p+Pb and p+p in future, will push our understanding of the dynamics of QGP medium.    

\section*{Acknowledgments}
RS acknowledges the support of the Polish National Science Centre grant 2019/35/O/ST2/00357 and 2023/51/B/ST2/01625 for the participation in the conference. RS also acknowledges the support of the organizers of ATHIC 2025. 
\bibliography{ref.bib}

\end{document}